\documentclass[traditabstract]{aa} 
\usepackage{graphicx}
\usepackage{txfonts}

\usepackage{natbib} 
\bibpunct{(}{)}{;}{a}{}{,} 
\usepackage{url}
\usepackage{multirow}

\usepackage{mydefinitions}

\begin{document}

\title{Chandra imaging of the $\sim$kpc extended outflow in 1H 0419-577}
\author{L. Di Gesu \inst{1}
    \and E. Costantini \inst{2}
      \and E. Piconcelli\inst{3}
     \and J.S. Kaastra\inst{2,4}
      \and M. Mehdipour \inst{2}
      \and S. Paltani \inst{1}}%
\institute{
Department of Astronomy, University of Geneva, 16 Ch. d’ Ecogia, 1290, 
Versoix, Switzerland \email{laura.digesu@unige. ch}
\and SRON Netherlands Institute for Space Research, Sorbonnelaan 2, 3584 CA Utrecht, The Netherlands 
\and Osservatorio Astronomico di Roma (INAF), Via Frascati 33, I--00040, Monteporzio Catone (Roma), Italy
\and Leiden Observatory, Leiden University, PO Box 9513, 2300 RA Leiden, the Netherlands}

\abstract{
The Seyfert 1 galaxy 1H 0419-577 hosts
a $\sim$kpc  extended outflow that is
evident in the [\ion{O}{iii}] image
and that is also detected as a warm 
absorber in the UV/X-ray spectrum.
Here, we analyze a $\sim$30 ks Chandra-ACIS
X-ray image, with the aim of resolving
the diffuse extranuclear X-ray
emission and of investigating
its relationship with the galactic
outflow. Thanks to its sub-arcsecond
spatial resolution,
Chandra resolves the circumnuclear X-ray
emission, which extends  up 
to a projected distance
of at least $\sim$16 kpc from the center. 
The morphology of the diffuse
X-ray emission is spherically symmetrical.
We could not recover a morphological
resemblance between the soft X-ray emission
and the ionization bicone that is traced
by the [\ion{O}{iii}] outflow.
Our spectral analysis indicates that
one of the possible explanations for the extended
emission is thermal emission from
a low-density 
($n_{\rm H} \sim 10^{-3} \rm \, cm^{-3}$) 
hot plasma ($T_{\rm e}\sim0.22$ keV).
If this is the case, we may be witnessing
the cooling of a shock-heated wind
bubble. In this scenario, 
the [\ion{O}{iii}] emission line
and the X-ray/UV absorption lines
may trace cooler clumps that are entrained
in the hot outflow. Alternatively, the extended
emission could be to due to a blend of emission lines
from a photoionized gas component
having a hydrogen column density of
$\nh \sim 2.1 \times 10^{22}$ \colc
and
an ionization parameter of $\log \xi \sim 1.3$. Because the source is viewed almost
edge-on
we argue that 
the photoionized gas nebula must be
distributed mostly along
the polar directions, outside
our line of sight. In this geometry,
the X-ray/UV warm absorber must trace
a different gas component, physically
disconnected from the emitting gas,
and located closer to the equatorial
plane.}
\titlerunning{}
\authorrunning{L. Di Gesu et al}
\maketitle
%
%

\section{Introduction}
%
%
%
%
%
\begin{table*}[ht]
\caption{Observation log.}     
\label{obs.tab}      
\centering                    
\begin{tabular}{l c c c c c}        
\hline\hline                 
Chandra sequence  & Observation ID  & Date & Net exposure 
& 0.2--1.0 keV counts \tablefootmark{*}
& 1.0--7.0 keV counts \tablefootmark{*}\\
 & & & ks & & \\
 \hline
 703141 & 18843 & 7 May 2016 & 8 & $283\pm17$  & $222\pm15$ \\
 703141 & 17464 & 8 May 2016 & 20 & $708\pm26$ & $561\pm24$\\
\hline 
\end{tabular}
\tablefoot{
\tablefoottext{*}{Net counts in an annulus comprised between 2\arcsec and 8\arcsec}}
\end{table*}
%
%
The X-ray emission 
from type-1 active galactic nuclei
(AGN) is dominated by the bright
nucleus. 
%
%
%
However, besides providing a diagnostic
of the physics in the vicinity of 
the black hole, the X-ray domain offers the possibility
of probing the extranuclear
AGN environment. For instance,
with the advent of the
Chandra satellite, X-ray imaging
at sub-arcsecond spatial resolution
became possible, thereby allowing detailed
studies of the circumnuclear gaseous 
environment in a handful
of nearby Seyferts.\\
Extranuclear X-ray emission
has been detected with Chandra
up to distances of the order
of  a few tenths of a kpc
both 
in Seyfert 2 (e.g.,
Circinus, \citealt{smi2001},
NGC 1068, \citealt{you2001},
NGC 4388, \citealt{cone4388})
and 
in Seyfert 1
galaxies (e.g.,
NGC 1365, \citealt{ngc1365wang2009},
NGC 4151, \citealt{wang4151a}).
In type-2 AGN it is often observed that
the X-ray morphology
resembles the ionization
bicone \citep{pog1988}, which is traced
for instance by the [\ion{O}{iii}] 
narrow line (NL).
The morphological
correspondence points to an origin in the same
medium photoionized by the central AGN
which has been, in
a few cases, confirmed using
photoionization analysis
techniques in combination
with high-resolution X-ray spectroscopy
\citep{bia2006,cone573}. However,
hot,
collisionally ionized plasma, may,
in some cases, coexist in the ionization cone
with the cooler photoionized
plasma and act as a confining
medium for the photoionized
clouds 
\citep[e.g., NGC 1365,][]{ngc1365wang2009}.\\

It has also been proposed
\citep{gua2007}
that the emitting gas
in the NL ionization cone
is the counterpart in emission
of the so-called warm absorbers
(WA) that are instead observed
in high-resolution X-ray/UV
spectra of nearby Seyfert-1
\citep[][]{cre2003,pic2005}.
These WA highlight gentle
outflows of photoionized gas
($v\sim$100--1000 \kms, \citealt{mck2007})
that may be located at distances
of the order 0.01--100 pc
\citep{cre2012} from the nucleus.
Spatially resolved spectroscopic
studies of the [\ion{O}{iii}] 
emission line
in nearby Seyferts
show indeed that the kinetics
of the gas in the ionization
cone is dominated by a radial outflow
\citep{fis2013},
which is another hint to
a possible connection. However,
proving a firm connection
would require
an accurate knowledge of
the physical parameters 
(the hydrogen column density \nh
and the ionization parameter
$\xi$) and
of both the emitting and the
absorbing gas, which is rarely
the case. 
Indeed, parameterizing
the NL emitting gas requires lengthy
photoionization calculations, that
have been carried out only in
a few cases so far
\citep[e.g.,                                                                                                                                ][] {nuc2010, dig2013,
whe2015, whe2016}.\\
The estimations \citep[e.g.,][]{det2010,ebr2016,
beh2017} of the kinetic luminosity
($L_{\rm kin}$)
carried by the WA in
Seyfert-type AGN indicates
that they are not energetically
significant for possible
AGN feedback \citep{dim2005,hop2010}. 
Galactic molecular winds, that are often
extended
on $\sim$kpc scales, 
and that can be either AGN 
\citep[e.g., Mrk 231,][]{fer2015}
or starburst
driven
\citep[see e.g.,][for a review]{vei2005},
may be a more effective
feedback agent
because they may be able
to displace 
a sizeable amount
of cold gas,
thereby leading to starvation of the star-formation
(SF) in the galaxy.
High spatial-resolution 
X-ray imaging is also
a way to test if and
how some feedback
is being exerted 
on the galactic medium,
because the mechanical shock
associated with
a large-scale outflow
is a viable mechanism
to heat the gas
up to X-ray energies.
Diffuse hot gas,
consistent with being
shock-heated by a starburst
wind, has been detected
on a 5 kpc scale in 
the Chandra image of
NGC 6240
\citep{ngc6240}.
In another work,
\citet{wan2010} discovered
diffuse soft X-ray emission in
the central $\sim$2 kpc
of NGC 4151, which can be
ascribed to a thermally emitting
gas, mechanically 
heated by an AGN
outflow.  A 65x50 kpc
X-ray nebula was also 
detected in a long-exposed
Chandra-ACIS image
of Mrk 231 
\citep{m231chandra}, which
is considered 
the prototypical
case of galactic-scale,
quasar-driven outflow
involving both
a neutral and a mildly
ionized phase of the medium
\citep{rup2013}.
In that case, the
lack of cool soft
X-ray emitting gas
in a region corresponding
to a large column
of outflowing neutral gas
can be interpreted
as evidence
of shock heating
due to the massive quasar 
outflow.\\ \\
%
%
The bright quasar 
known as 1H 0419-577 
(or IRAS F04250-5718) 
is spectrally classified 
as Seyfert 1.5 
\citep{ver2006}, and
is a radio-quiet AGN located
at a redshift of 0.104.\\
%
%
It was previously found that,
at X-ray energies smaller than
$\sim$2 keV,
this source displays a highly variable
soft-excess \citep{sin1985}, with frequent transitions
between low and high flux states
on timescales of months/years \citep{gua1998}.
The quest of understanding the origin
of this variability motivated
a rich, multi-epoch
X-ray spectral coverage
that includes six XMM-Newton
\citep{pag2002,pou2004a,pou2004b}
and two Suzaku observations
\citep{tur2009,pal2013}.
%
%
%
In \citet{dig2014}, 
we fitted the deep (97 ks)
XMM-Newton EPIC-pn spectrum
together
with the optical/UV flux data points
with a Comptonization 
model.
In this model, the optical/UV
disk photons are up-scattered
to X-ray energies by two
intervening layers
of plasma, with $T\sim100$ keV
and $T\sim0.5$ keV,  respectively.
With the addition
of an intervening neutral absorber,
variable in column density, our model
is able to explain also the historical
optical-to-X-ray spectral variability
of the source.\\ 
This long XMM-Newton observation
was taken simultaneously with
a HST-COS observation in the UV,
with the aim of detecting
and characterizing a photoionized
WA. In the high-quality
RGS spectrum we detected
the absorption lines
(e.g., \ion{O}{iv}--\ion{O}{vi})
of a lowly ionized
WA ($\log \nh \sim 19.9$ \colc,
$\log \xi \sim 0.03$ \xiun), 
which is consistent
with being one and the same
as the UV absorber.
Combining X-ray and UV diagnostics,
we estimated that the absorption
lines arise in a galactic
wind, located at a distance
of at least $\sim$4 kpc from the nucleus.
Since then, the estimated
location was confirmed
using Integral Field
Unit (IFU) optical spectroscopy
\citep[at the GEMINI telescope][]{liu2015}.
The velocity-resolved [\ion{O}{iii}]
map shows indeed 
a biconical outflow
extending up at to at least
$\sim$3 kpc from center. The outflow
axis is tilted at $\sim$20\degr with respect
to the galactic disk and the opening
angle of each cone is $\sim$70\degr.\\
Extended, circumnuclear X-ray
emission from 1H 0419-577
was noticed for the first time
in the ROSAT HRI image. After
a careful recalibration
of the HRI PSF, which was carried
out using a sample of bright
point-like sources 
\footnote{see the ROSAT calibration document
in the MPA website 
(\url{http://www.mpe.mpg.de/xray/wave/rosat/calibration/hri/psf/hri_psf.php})}, 
\citet{pre2001}
concluded that \virg{the Seyfert
1 galaxy 1H 0419-577 seems to be
really extended}.\\
Here, we present the
results of a Chandra-ACIS
observation of 1H 0419-577,
that we performed with the aim
of resolving
the
extended X-ray emission
in this source
and of investigating its
possible connection
with the galactic outflow.
%
%
%
The plan of the paper
is as follows. In Sect.
\ref{obs} we describe the data preparation
while in Sects. \ref{img}
and \ref{spec} we present the imaging
and the spectral analysis, respectively.
Finally, in Sect. \ref{disc} we discuss
our results and in Sect. \ref{sum} we present
our conclusions.\\
For the present 
analysis, we considered a flat universe, with
\ho=70 \kmsmpc, \omegam=0.3, \omegalambda=0.7.
With these cosmological parameteres, 
1\arcsec corresponds
to 1.9 kpc at the redshift
of 1H 0419-577.
The C-statistics \citep{cas1979}
is used throughout the paper
and errors are quoted
for $\Delta C=2.7$. In all the
spectral models presented in the
following we consider the Galactic 
hydrogen column density
from \citet[][\nh=$1.26 \times 10^{20}$ \colc]{kal2005}.
 \begin{figure*}[t]
 \centering
 \includegraphics[width=1.0\textwidth]{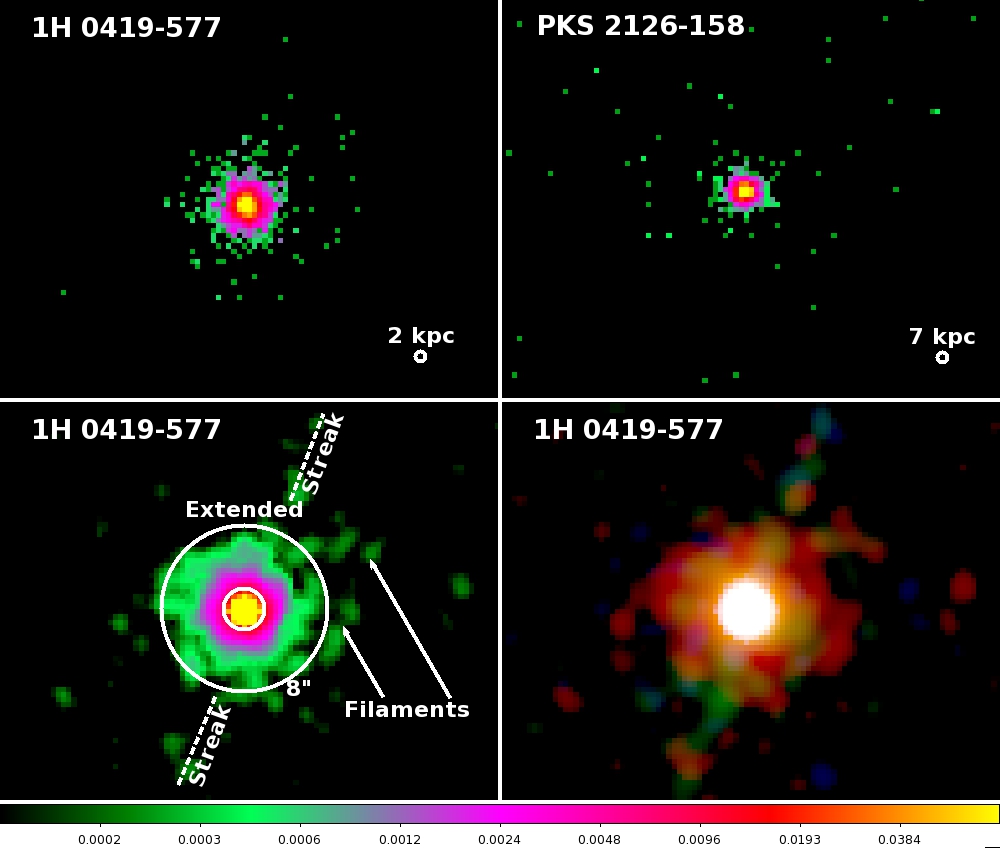}
    \caption{Chandra-ACIS images of
    1H 0419-577 (\textit{top left}) and of the point-like source PKS 2126-158 (\textit{top right})
    in the 0.2--1.0 keV band. As a visual aid, we also show the color bar of these two images
    at the bottom. For comparison, 
    we show in both the panels a circle with the radius of the core of the Chandra PSF. \textit{Bottom left}: 
    Same as first panel, smoothed using a Gaussian kernel of 3 pixels.
    The regions that we used for spectral extraction are overlaid. We also mark the direction
    from which we removed  the ACIS readout streak (dashed line) and the location of possible extended 
    filaments around the source. 
    \textit{Bottom right}: False-color image obtained using the PSF-deconvolved 
    0.2--0.5, 0.5--2.0, and 2.0--7.0 keV band images as red, green, and blue, respectively.}
    \label{img.fig}
 \end{figure*}
%
%
%
\section{Observations and data preparation}
\label{obs}
\subsection{Data reduction}
In Table \ref{obs.tab} we outline the summary
of our Chandra observation of 1H 0419-577 which
was carried out in May 2016. The total observing time of $\sim$ 30 ks
was split into two shorter exposures of $\sim$8 and $\sim$20 ks, 
respectively.
Our observation was performed with only the back-illuminated
S3 chip of the ACIS detector on. We chose the $1/8$ subarray
configuration, which reduces the frame time to $\sim$0.4 s,
to minimize a possible nuclear pile-up. \\
Throughout the paper, we compare 1H 0419-577
with PKS 2126-158, a bright blazar located at z=3.26,
which can be considered a prototypical point-like
source. We retrieved the Chandra data of PKS 2126-158,
that were taken with the same detector configuration
as our observation,
from the public Chandra archive (Obs ID: 376).\\
We performed the data analysis using 
the latest version of the software
CIAO (v. 4.9) in combination
with the most up-to-date calibration database CALDB 4.7.3.
We reduced all the Chandra data using the
chandra\_repro script and 
we removed the
ACIS readout-streak from the calibrated event-files.
For 1H 0419-577, we
created a merged, $\sim 28$ ks exposed event-file.
For this purpose, we first cross-matched the WCS 
coordinates of the two event files using
those of the longest one as reference. We ran
the wavdetect task for source detection
for both the event files,
and we input the two obtained 
source lists in the 
CIAO task reproject\_aspect.
We set the task to
apply
the WCS correction to the 8 ks event-file,
which minimizes the differences
between the two source lists.
The source centroid
found by wavdetect
is 4:26:00.7, -57:12:02,
which is consistent
with the nominal
coordinates of 1H 0419-577.
We use these coordinates
in the remainder of the analysis.
Hence,
we merged the corrected 8 ks event-file
and the 20 ks event-file using
the CIAO tool reproject\_obs.\\
We checked the event-files
for possible photon pile-up.
The CIAO task  pileup\_map 
creates an image in
counts per ACIS frame,
which can be used to estimate
the pile-up fraction. The map
shows that the nucleus of 1H-0419-577
is piled-up at a level of
$\sim$10\% in a 3x3
pixels island around
the nucleus. We consider
this level of pile-up
acceptable for our analysis
and we comment on it
below. Finally,
we note that in the observation
of PKS 2126-128, that we use
as a comparison, the pile-up fraction
is the same.\\
\subsection{PSF simulations}

In order to disentangle 
possible extended emission
from the bright point-like
nucleus,
accurate modeling of the
instrumental point spread
function (PSF) is needed.
We used the Chandra HRMA 
ray tracer (ChaRT) to simulate 
a Chandra observation
of a point-like source having
the same spectrum as the
nucleus of 1H 0419-577.
We extracted the nuclear
spectrum in a circular region
having a radius of 2\arcsec 
from which we excluded
a 3x3pixels box including
the piled-up area.
A phenomenological
model, comprising a power-law
plus a blackbody provides
an adequate fit
spectrum of the nucleus
extracted in this region.
Besides the
model spectrum of the nucleus,
we input in ChaRT the same satellite configuration 
of our observation and the
aspect-solution of the 20 ks event file.
We treated the ChaRT
outputs with the CIAO script  
simulate\_psf, which uses MARX (v. 5.3.0) 
to project rays onto the detector and 
to create the PSF pseudo
event-file. Starting
from that, we created
images and surface-brightness profiles 
of the PSF
in the 0.2--1.0 keV
and 1.0--7.0 keV
energy bands. We used
the PSF images (see below)
for the PSF deconvolution
of the source images that
we performed using
the Lucy-Richardson
\citep{lucy1974}
method as implemented
in the arestore task
in CIAO.

\subsection{Spectra preparation}
\label{specprep}

We extracted all the spectra and spectral
response matrices using the script specextract,
which is provided in CIAO. We took the
spectrum of the nucleus from a circular
region with a radius of 2\arcsec.
Conversely, we extracted the spectrum of the
extended emission in an annulus centered
on the same coordinates and with 
an inner and an outer radius of 2\arcsec 
and 8\arcsec, respectively. As
a comparison (see below),
we
also extracted the spectra of PKS 2126-128
in the same regions. 
In all cases, 
we took the
spectrum of the background from 
an exterior annulus
3\arcsec wide.
%
%
%
For 1H 0419-577, we extracted the
spectra from each individual observation
as recommended in the CIAO guidelines.
Hence, in all cases,
we co-added the spectra in a single
28 ks exposed spectrum. We summed
the spectra using the FTOOL
mathpha and averaged the ancillary
response files with addarf.\\
All the spectral
analysis outlined in Sect. \ref{spec}
was performed using
the latest
version of the SPEX program
(v.3.03.00). The C-statistics,
its expected value, and its variance
are computed according to the
equations given in \citet{kaa2017}.
In all the fits, 
we accounted for the cosmological redshift of the source
and the Galactic absorption. For the
latter, we used a collisionally ionized
absorption model in SPEX (HOT)
where we kept the temperature
fixed at 0.5 eV to mimic
a neutral gas.
%
\begin{table}[t]
\caption{Comparison between the soft X-ray surface brightness
inside and outside the [\ion{O}{iii}] outflow regions.}     
\label{test_ref.tab}      
\centering                    
\begin{tabular}{l c c }        
\hline\hline                 
Region 
& P.A.\tablefootmark{a} 
& 0.2--1.0 keV surface brightness \tablefootmark{b} \\
 & & $\rm counts/s/pixel^2$\\
 \hline
\textit{no-outflow 1} & -10\degr--20\degr & $1.7\pm0.2$\\
\textit{outflow 1} & 20\degr--180\degr& $2.51\pm0.02$ \\
\textit{no-outflow 2} & 180\degr--210\degr& $1.4\pm0.4$ \\
\textit{outflow 2} & 210\degr--350\degr& $2.51\pm0.08$\\
\hline
\end{tabular}
\tablefoot{
\tablefoottext{a} {Position angle of the pie-shaped regions, as shown in Fig. \ref{o3.fig}, left panel.}
\tablefoottext{b}{Background subtracted surface brightness after PSF deconvolution.}}
\end{table}
%

 \begin{figure}[t]
 \centering
 \includegraphics[width=0.5\textwidth]{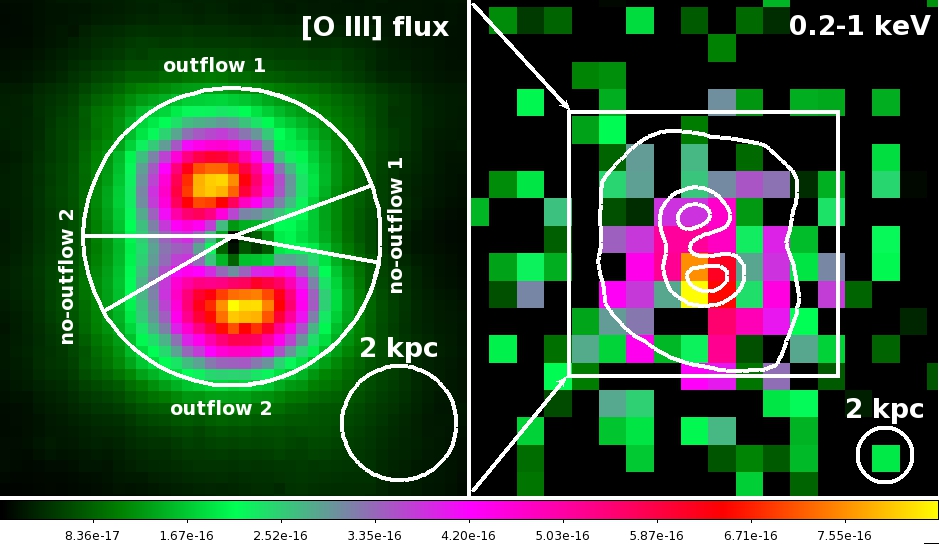}
    \caption{Comparison between the map of 
    the [\ion{O}{iii}] surface brightness (\textit{left panel}, courtesy of Guilin Liu) 
    and the PSF deconvolved, 
    unsmoothed, 0.2--1.0 keV image (\textit{right panel}). For comparison, a circle with the
    radius of the core of the Chandra PSF is shown in both panels.  In the left panel we show
    the region that we used to test the symmetry of the soft X-ray extended emission (see the text for details).
    The color bar of the [\ion{O}{iii}] surface-brightness map is also shown.
    In the right panel we overlay the [\ion{O}{iii}] contours 
    on the soft X-ray image. The box represents the area covered 
    by the GEMINI field of view. }
    \label{o3.fig}
 \end{figure}


\begin{figure*}[t]
\begin{minipage}[c]{1.0\textwidth}
 \includegraphics[width=0.5\textwidth]{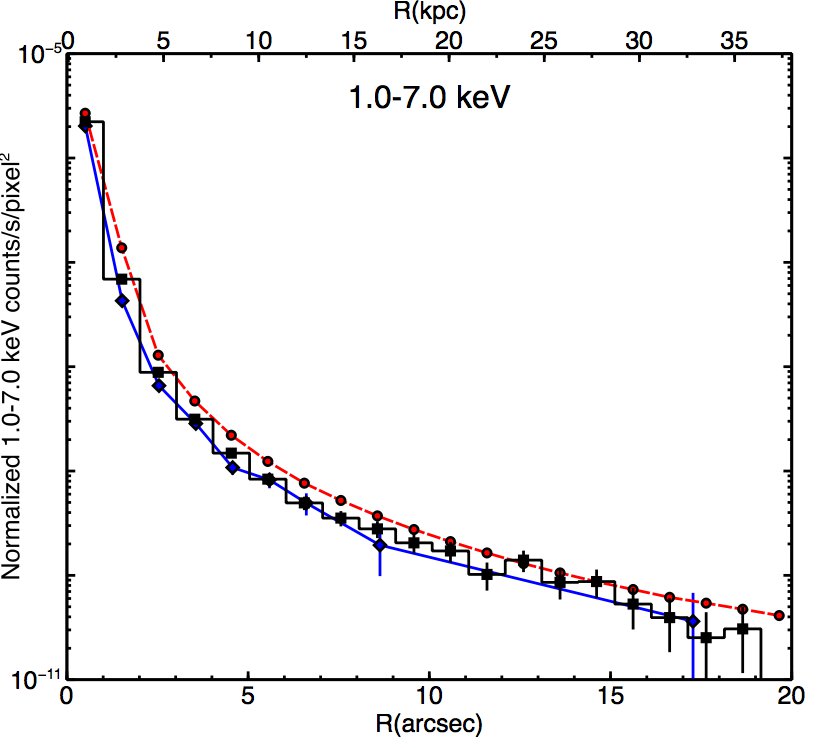}
 \hspace{0.3cm}
 \includegraphics[width=0.5\textwidth]{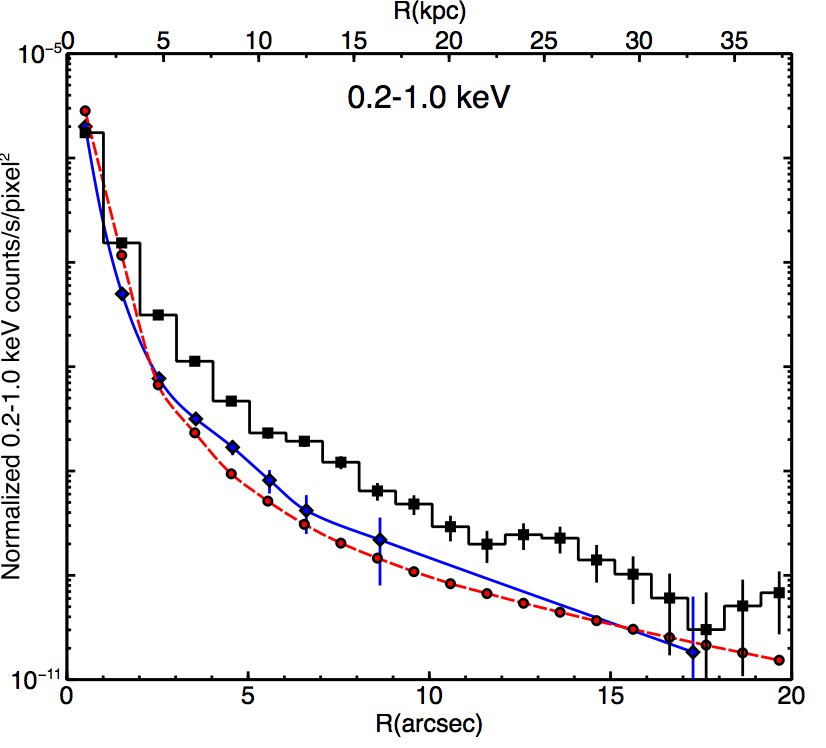}
\end{minipage}
  \caption{Comparison among  the surface-brightness
  profiles of 1H 0419-577 (histogram, filled squares),
  those of PKS 2126-158 (solid line, filled diamonds),
  and those of the Chandra PSF (dashed line, filled circles).
  We show the comparison in the 1.0--7.0 keV (\textit{left panel}) and in the
  0.2-1.0 keV (\textit{right panel}) energy bands. All the profiles are background
  subtracted and normalized to one. We note the logarithmic scale
  on the vertical axes.}
  \label{sbr.fig}
\end{figure*}
%

\section{Imaging analysis}
\label{img}

\subsection{Images}
In Fig. \ref{img.fig}
we compare the morphology
of 1H 0419-577 (top left)
with that of the point-like
source PKS 2126-158 (top right)
in the 0.2--1.0 keV band. To help
the comparison, we normalized
both the images to the maximum
and we matched the color
scales. In this band, 1H 0419-577
clearly displays broader
wings than those of
a point-like source.
In order to highlight the
soft X-ray morphology, we also show
the smoothed 0.2-1.0 keV image
of 1H 0419-577 (bottom right) 
and the false color image (bottom left).
The bright nucleus 
of 1H 0419-577 
is surrounded
by a halo of soft X-ray emitting
material which extends up to a distance of
at least $\sim$ 8\arcsec/16 kpc and 
has  a spheroidal morphology. 
We note that in the west direction,
the soft X-ray emission
could be extended even
beyond this distance
since hints of faint filamentary
structure (Fig. \ref {img.fig}, bottom left panel) 
are seen up to distances
of $\sim$30 kpc (see below).
These features 
cannot be due
to an imperfect removal of the
ACIS readout streak,
as we show in the bottom-left panel of Fig. \ref{img.fig}. Therefore, they are likely
associated to the extended emission of the source.
However, because
of the low signal-to-noise ratio
we did not consider these features
in the present analysis.\\
After having established
that the source displays
extended features
for energies below 1 keV,
it is instructive to compare 
the morphology of the soft X-ray emission
with that of the [\ion{O}{iii}]
emission line highlighted
in the GEMINI-IFU image.
A morphological resemblance between
the NLR traced by [\ion{O}{iii}]]
and the extended soft X-ray
emission is often
interpreted as an indication
that these two emission components
arise in the same medium photoionized
by the central AGN \citep{bia2006}.
In the PSF-subtracted
[\ion{O}{iii}] surface brightness map
(Fig. \ref{o3.fig}, left panel,
courtesy of Guilin Liu)
the source displays the typical
morphology of an edge-on disk galaxy.
Two [\ion{O}{iii}]-bright blobs,
are clearly seen
above and below a roughly vertical gap
that traces the obscuring 
disk material. In the analysis
performed in \citet{liu2015}, it is
also found that the [\ion{O}{iii}]
emitting material in the blobs
is outflowing at velocities
of the order of hundreds of \kms.
Moreover, they set a lower limit of 
$\sim$1.5\arcsec/3kpc
for the maximum extension
of the outflow. They note,
however, that the [\ion{O}{iii}]
line is strongly detected in the
entire 10x10 kpc$^2$ GEMINI 
field of view.\\ 
The size of the
GEMINI field of view
is smaller than the
size of the soft X-ray
halo, which extends
up at least $\sim$16 kpc.
In the right-hand panel of Fig. \ref{o3.fig}, we show
a zoom-in of the 0.2--1.0 keV image in the
area where the [\ion{O}{iii}] bicone
(overlaid contours) is seen.
Since the size
of the [\ion{O}{iii}] lobes
is comparable with the size
of the core of the Chandra-PSF
we used the PSF deconvolved
image to perform this
comparison. Indeed,
removing the effect of PSF
could  be helpful to recover 
a possible 
[\ion{O}{iii}]/soft X-ray morphological
resemblance in type-1 AGN, where
the nuclear light outshines
the extended emission
\citep{gom2017}.
Here we found that,
after the PSF deconvolution,
we are unable to recover the [\ion{O}{iii}] 
biconical morphology 
in the soft X-ray image.
The gap that separates
the two [\ion{O}{iii}]-blobs
is not present in the 0.2--1.0 keV band,
where the morphology is spherically
symmetric, rather than biconical. \\
In order to quantitively test
this judgement, we used the deconvolved
image to compute  the soft X-ray surface brightness
in the regions where the [\ion{O}{iii}]
outflow is seen and we compare it
with that measured in the regions
external to the outflow. For this exercise,
we used four pie-shaped regions covering
the 0"--8" radial range. The position angles
(P.A.) of each sector were chosen using
the [\ion{O}{iii}] surface brightness
map as a guide (see Fig. \ref{o3.fig}, left panel). The results
are outlined in Table \ref{test_ref.tab}.
We found that the difference
in  surface brightness between
the \virg{\textit{outflow}} and
\virg{\textit{no-outflow}} regions
has a low statistical
significance (i.e.,$< 3\sigma$).
Thus, this quantitive test indicates
that indeed the soft 
X-ray extended emission may not show
the same asymmetry that is clearly seen
in the [\ion{O}{iii}]
image.

\subsection{Surface-brightness profiles}
\label{souprof.ss}

We extracted the surface-brightness profiles
of the source in the 0.2--1.0 keV and
1.0--7.0 keV bands. For this exercise,
we used 20 concentric
circular annuli, each with a
width of 1\arcsec . 
Conversely, we extracted
the surface brightness 
of the sky background using the
same region of Sect. \ref{specprep}.
As a comparison, we extracted
also the surface-brightness
profiles of the simulated PSF
and of the point-like source 
PKS 2126-158.
To help the comparison,
we normalized all the
profiles to one.\\
We show all the surface-brightness profiles 
in Fig. \ref{sbr.fig}.
The mismatch between
the profile of source (histogram, filled squares) 
and that of the simulated
PSF (dashed line, filled circles) 
in the first annulus 
is due to the nuclear
pile-up, which artificially
lowers the number of counts
at low energies. Indeed,
the peak of the surface-brightness
profile of 1H 0149-577 is consistent
with that of PKS 2126-528 (solid line, filled diamonds)
which has a similar pile-up fraction.
Thus, in our comparisons,
the profile of PKS 2126-528
conveniently illustrates 
how a pile-up of the
order of 10\%
modifies the PSF
in the first 2\arcsec.
At larger radii, the pile-up
has no effects 
\citep[see e.g., NGC 4151,][]{wang4151a}.
The small discrepancy
between the profiles of PKS 2126-5128
and those of the simulated PSF could be due
to changes of the Chandra-PSF since
the time the source was observed.
(i.e., 1999). Indeed, the build-up
of material on the optical blocking 
filter is known to affect the PSF.
\\
The profile of  Fig. \ref{sbr.fig},
left panel,
indicates that at energies
above 1 keV the source
is consistent with being
point-like.
On the other hand,
at energies below 1 keV (right panel),
the surface-brightness profile
of 1H 0419-577 differs significantly
from what is
expected from an unresolved
point-like source.
The profile indicates
that, at these energies,
the extended
emission of the source
is resolved by Chandra.
The soft X-ray emission
of 1H 019-577
appears to be extended
up to a distance of
15\arcsec/30 kpc
from the center. 
As we showed in
Fig. \ref{img.fig},
the bulk of the
extended emission
comes from a spheroid
which extends up to
a distance of
8\arcsec/16 kpc.
However, beyond this distance,
hints of faint filamentary
features are seen. These
are likely dominating
in the tail 
of the surface-brightness
profile.

\section{Spectral analysis}
\label{spec}

%
\begin{table}
\caption{Best-fit parameters for the spectrum of the nucleus. 
Values without errors were kept frozen in the fit.}     
\label{fitnuc.tab}      
\centering                    
\begin{tabular}{l c c c}        
\hline\hline                 
Model component & Parameter & Value & Units \\
\hline
\multirow {3}{*}{Warm COMT}&  
$\tau$ \tablefootmark{a} & 7 &  \\
& $T$\tablefootmark{b} & 0.7 & keV\\
& $F^{\rm wc}_{\rm 0.5-10.0\, keV}$ \tablefootmark{c}  
& $0.9\pm0.8$& $10^{-12}$ \ergsc \\
\hline 
 \multirow {3}{*}{Hot COMT}&  
 $\tau$\tablefootmark{a} & 0.6 &  \\
& $T$ \tablefootmark{b}& 150 & keV\\
& $F^{\rm wc}_{\rm 0.5-10.0\, keV}$ \tablefootmark{c} 
&$11.1\pm 0.2$ & $10^{-12}$ \ergsc \\
\hline
 \multirow{2}{*}{Neutral absorber} 
 & \nh \tablefootmark{d} & $1.7\pm{0.7}$ & $10^{22}$ \colc \\
& $C_{\rm V}$ \tablefootmark{e} &$0.16^{+0.18}_{-0.08}$ & \\
\hline 
\end{tabular}
\tablefoot{
\tablefoottext{a}{Plasma optical depth.}
\tablefoottext{b}{Plasma temperature.}
\tablefoottext{c}{Observed flux in the quoted band.}
\tablefoottext{d}{Hydrogen column density.}
\tablefoottext{d}{Covering fraction.}}
\end{table}

%

 \begin{figure}[t]
 \centering
 \includegraphics[width=0.5\textwidth]{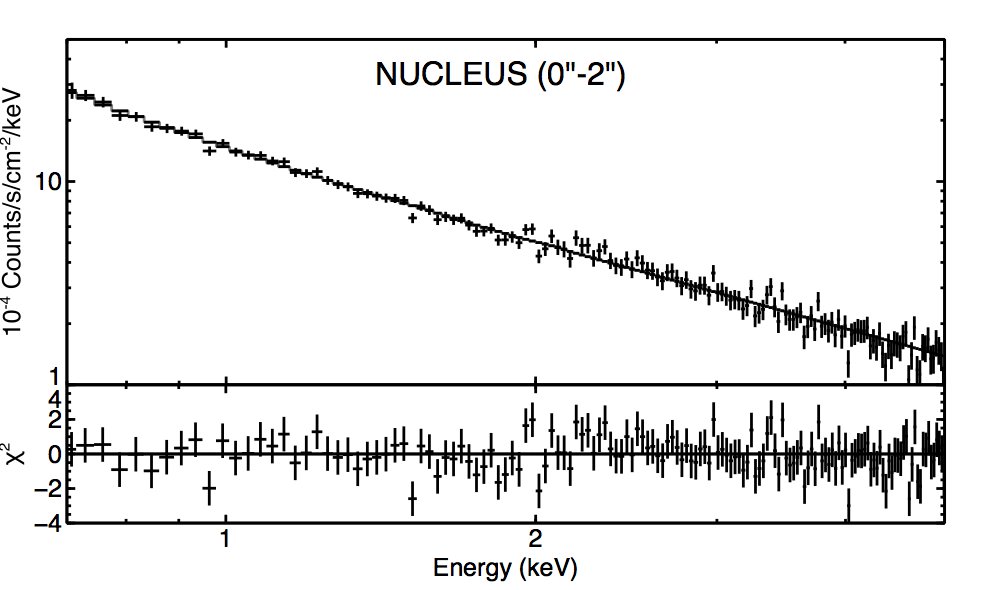}
    \caption{Best-fit for the spectrum of the nucleus 
    (\textit{top panel}) and spectral
    residuals in terms of $\chi^2$(\textit{bottom panel}).
       The spectrum is shown in the 0.7--5.0 keV band.
        The solid line represents our best-fit Comptonization model. }
       \label{fitnuc.fig}
 \end{figure}

\subsection{The spectrum of the nucleus}
\label{nspec.ss}

In the context of this work, a reliable modeling
of the nuclear spectrum is mainly needed
to quantify how much the nuclear
light scattered by the PSF wings contaminates
the spectrum of the extended emission at radii
above 2\arcsec. For this purpose, we assume
the variably absorbed Comptonization model
of \citet{dig2014}.
Here, we fitted the spectrum of the nucleus in the
0.7--5.0 keV band. 
We used two Comptonized components
(COMT model in SPEX \citealt{tit1994}) where
we  kept the plasma parameters
(temperature $T$ and optical depth $\tau$)
frozen to the average of the values observed
in the XMM-Newton dataset. Hence, we let the
normalizations free to vary. This fit
is already statistically acceptable
with C/Expected-C=141/148. However,
we checked whether the addition
of a neutral absorber could
improve the fit. Indeed, the
trend between absorbing column density
and observed flux that we noticed
for the XMM-Newton dataset would predict
a column-density of the order of a few
$10^{22}$ \colc  at the flux level
of the Chandra observation 
($\sim8.4 \times 10^{-12}$ \ergsc),
which is a low flux state for this
source. By adding to the fit
a neutral absorber with free
column density and covering fraction,
we obtained a marginally significant
improvement of the fit ($\Delta C=-8$).
However, since the obtained parameters of the absorber
are in line with the expectation
for this flux level, 
we adopted this case as our
best fit for the nuclear spectrum. We show
the best-fit model for the
spectrum of the nucleus in
Fig. \ref{fitnuc.fig} and we
outline the final best-fit parameters 
in Table \ref{fitnuc.tab}.

\subsection{Modeling the nuclear contamination at r$>$2\arcsec}

\begin{figure}[t]
 \centering
 \includegraphics[width=0.5\textwidth]{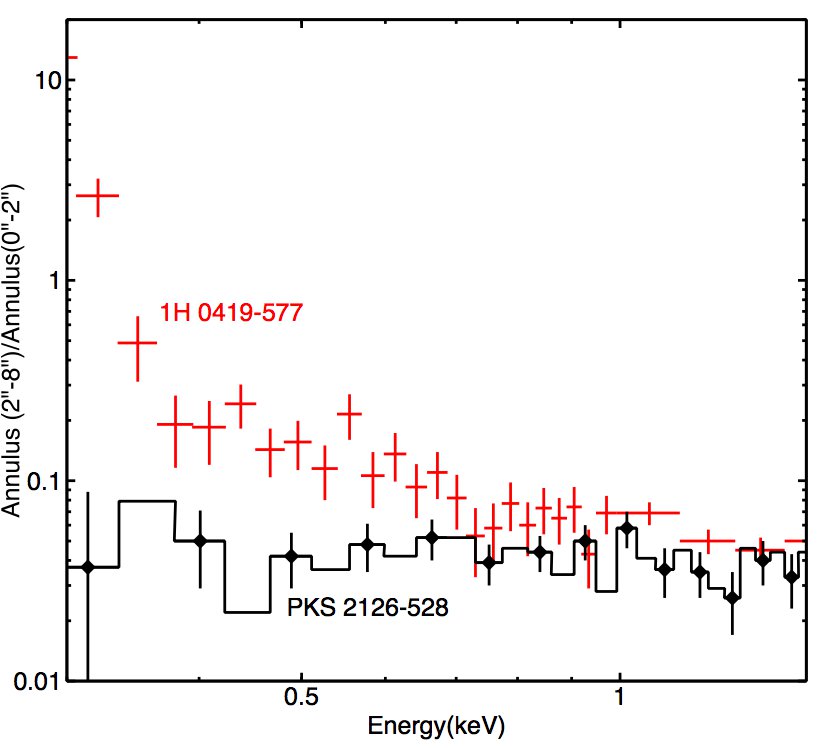}
    \caption{Ratios between the spectrum taken between 0\arcsec and 2\arcsec
    and between 2\arcsec and 8\arcsec from the center, for 1H 0419-577
    and the point-like source PKS 2126-528.}
       \label{specratio.fig}
 \end{figure}

The spectrum of the nucleus is scattered
by the wings of the instrumental PSF
at radii larger than the PSF core. As
a consequence, any spectrum taken off-nucleus 
will be slightly contaminated by this
scattered light. Here we show,
in a model-independent way, that
the nuclear spectrum scattered by the
PSF wings is not sufficient to explain the
spectrum of the extended emission
of 1H 0419-577.\\
In Fig. \ref{specratio.fig}
we plot the ratio between the spectrum
taken in the PSF-wing region (between 2\arcsec and 8\arcsec) and
in the PSF-core region (between 0\arcsec and 2\arcsec) for
1H 0419-577 and for the point-like
source PKS 2126-528. We use
the 0.3--2.0 keV band to perform
this comparison. For a point-like
source such as PKS 2126-528
(histogram), this ratio is roughly
a constant because it is,
at any energy, simply equal
to the instrumental ratio between
the PSF wings and the
PSF core 
(i.e., $wings\_to\_core\sim$0.04, see Fig. \ref{specratio.fig}).
Conversely, for 1H 0419-577,
the ratio steepens up at energy
below $\sim$1.0 keV, significantly exceeding
what is expected from the
scattering at larger radii
of the spectrum
of the central point-like
source. This confirms that
an extended emission
component is present
in the spectrum of 1H 0419-577. In
all of the following 
analyses,
we modeled the contribution
of the scattered nuclear
light in the spectrum of the extended emission
using the model of the nuclear
spectrum of Sect. \ref{nspec.ss}, 
with the normalizations
rescaled by the factor
$wings\_to\_core$.

\subsection{The spectrum of the extended emission}

\begin{figure*}[t]
\begin{minipage}[c]{1.0\textwidth}
 \includegraphics[width=0.5\textwidth]{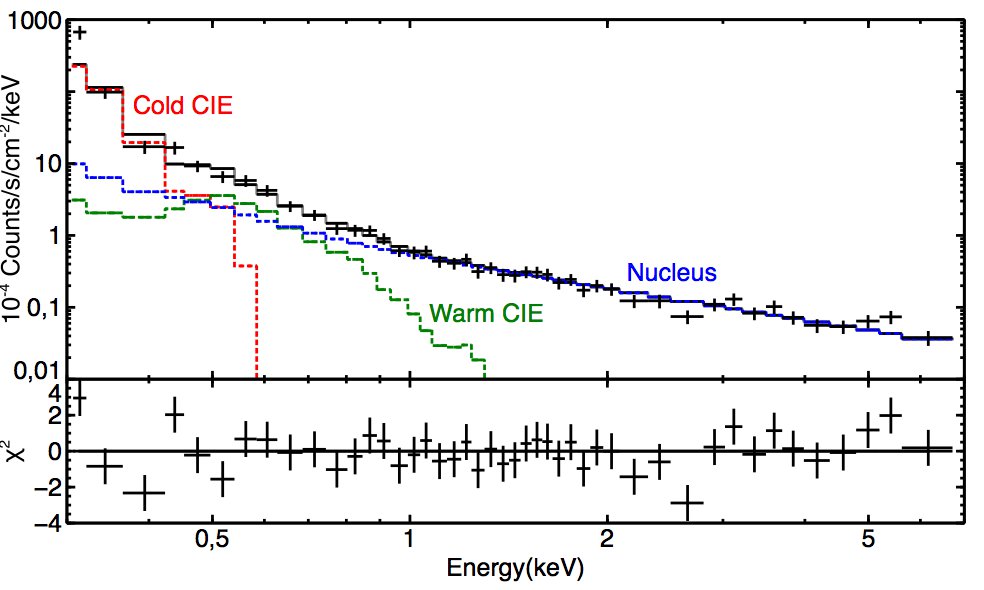}
 \hspace{0.3cm}
 \includegraphics[width=0.5\textwidth]{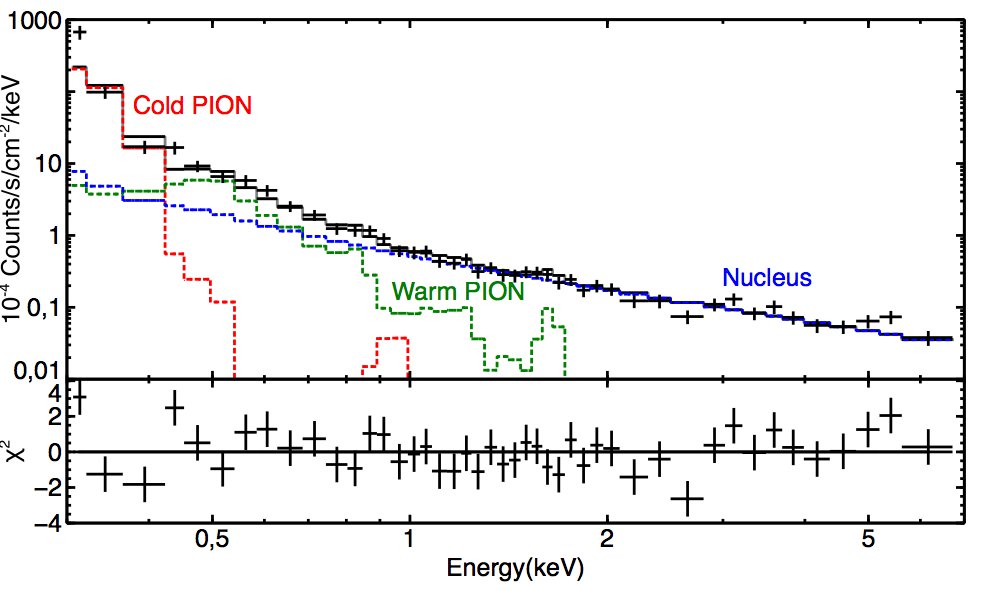}
\end{minipage}
  \caption{Best-fits of the spectrum of the extended emission
  using a collisionally ionized plasma model (\textit{left panel})
  and
  a photoionized plasma model (\textit{right panel}).
  In each panel, we show the total best-fit
  model (solid line), the spectral components
  separately (dotted lines),
   and the residuals in terms
  of $\chi^2$. The spectral components
  are labeled as in Table \ref{fitext.tab}.}
  \label{extspec.fig}
\end{figure*}

We fitted the spectrum of the extended emission
in the 0.3--7.0 keV band. A model including
only the rescaled nuclear spectrum adequately fits
the data only at energies above 1 keV 
(Fig. \ref{extspec.fig}),
where indeed the source
is unresolved (Sect. \ref{souprof.ss}). The C-statistics 
for this model over the 0.3--7.0 keV
band is poor (C/Expected-C=430/45), which
confirms a need for some additional
extended component at low-energies.\\
At first, we tested whether
the spectrum could be due
to the thermal emission
from a hot, collisionally ionized plasma.
For this,
we used the CIE model in SPEX
which is based on the plasma calculations
of \citet{kaa1996}. The free parameters
of the model are the electron temperature
($T_{\rm e}$) and the emission measure of the plasma
$Norm$. We consider an isothermal plasma
with zero turbulent velocity. The default
metal abundances in SPEX are protosolar abundances
from \citet{lod2009}.\\
The addition of a CIE component with $T_{\rm e}\sim 0.22$ 
keV (hereafter warm CIE) resulted in a statistically
significant improvement of the goodness of the fit.
However, the C-statistic for this model
(C/Expected-C=191/45) remains poor. Indeed,
the spectrum rises up sharply towards lower energies, significantly
exceeding what is expected from a single-temperature
plasma (Fig. \ref{extspec.fig}, left panel). 
Thus, we added another CIE
component (hereafter cold CIE) to the fit. We found
that the spectrum in the 0.3--7.0
keV band is best-fitted (C/Expected-C=56/45)
by the combination of a warm and
cold ($T_{\rm e} \sim 0.02$ keV) CIE
component, besides the contribution
of the nuclear spectrum
scattered by the PSF wings.
The decrease of the C-statistic due
to the addition of the cold CIE
component is $\Delta C$=-135.
In Table \ref{fitext.tab}, upper panel, we
list all the parameters and the errors for
this model.\\
Photoionization by the central
AGN is another possible physical scenario for
the extended emission in Seyfert
galaxies. The latest version of SPEX includes
for the first time a model for a photoionized
plasma in emission. In this PION
\footnote{see the SPEX manual: \url{http://var.sron.nl/SPEX-doc/manualv3.02/manual.html}} model
\citep{pion},
the photoionization equilibrium
of the plasma is computed self-consistently
using the same continuum of the fit
as ionizing SED. In our case, 
we used the rescaled nuclear continuum;
the  PION ionizing continuum.
The free parameters of this
model are the hydrogen column density
of the plasma \nh and the ionization
parameter $\log \xi$. We noted
that, in our fits,
\nh is degenerate with the
$\Omega$ parameter in PION,
that is a scaling factor related
to the geometry of the emitting plasma.
Thus, for practical reasons,
we considered a spherical shell
of material covering the entire  
sky and a low optical depth
plasma out of which both forward
and backward radiations can escape
($\Omega$=2, $mix=0.5$).
We considered only the emission
from the plasma, thus we set
a null covering fraction
in absorption.\\ 
At first we tested if
line-emission from a plasma
with the same parameters
($\log \nh=19.9$ \colc,
$\log \xi=0.03$ \xiun)
of the known WA could account
for the observed extended
emission. We found that,
in the assumed geometry, 
the emission
from the WA is too weak
to account for extended
emission. The spectrum
is poorly fitted (C=438/66) and large
residuals are left
below 1 keV. From this
exercise, we conclude
that the counterpart
in emission of the WA
can account at most
for 2\% of the
0.5--2.0 keV luminosity
of the extended emission.\\
Thus,
after having established
that plasma parameters different
from those of the WA are needed,
we let \nh and $\log \xi$ free
to vary in the fit.
The behavior of this photoionized
plasma model (Fig. \ref{extspec.fig}, right panel)
is similar to what we described
above for the case of a collisionally ionized
plasma model. A single PION component
with $\log \xi=1.29$ (hereafter warm PION) 
is unable to model the
steepening-up of the
spectrum at low-energies
(C/Expected-C=206/43).
We obtained the best fit
(C/Expected-C=72/43) by
adding a second PION component
with a lower ionization parameter
($\log \xi=-4$, hereafter cold PION). 
We outline the parameters and the
errors for the final best-fit
model in Table \ref{fitext.tab}, 
lower panel.\\
In conclusion, this fitting
exercise indicates that
the extended emission
in 1H 0419-577 is consistent
with both thermal emission
from a hot plasma and
with photoionized emission
from a warm plasma.
Although the C-statistic
shows a preference for
a collisionally ionized plasma
model, a photoionized plasma
model cannot be ruled
out. 
In the following,
we discuss the physical
plausibility of both
scenarios, 
considering the large scale outflow
that is known for this
AGN.

%
%
%
%
%

%
\begin{table}
\caption{Parameters for two for the spectrum of the extended emission. }     
\label{fitext.tab}      
\centering                    
\begin{tabular}{l c c c c }        
\hline\hline 
\multicolumn{5}{c}{Collisionally ionized plasma model}\\
Component & Parameter & Value & Units & $\Delta C$ \tablefootmark{a} \\
\hline
\multirow{2}{*}{Warm CIE}&  $T_{\rm e}$ \tablefootmark{b} & $0.22\pm0.02$ & keV 
& \multirow{2}{*}{-239}  \\
& $Norm$  & $2.8\pm0.4$ & $10^{55} \, \rm cm^{-3}$ &\\
\hline
\multirow{2}{*}{Cold CIE}&  $T_{\rm e}$ \tablefootmark{b} & $0.020\pm0.002$ & keV 
& \multirow{2}{*}{-374}  \\
& $Norm$  & $7^{+0.2}_{-2} \times 10^{7}$ & $10^{55} \, \rm cm^{-3}$ &\\
\hline \hline
\multicolumn{5}{c}{Photoionized plasma model}\\
Component & Parameter & Value & Units & $\Delta C$ \\
\hline
\multirow{2}{*}{Warm PION}&  $\log \xi $ \tablefootmark{c} & $1.3\pm 0.1$ &  \xiun 
& \multirow{2}{*}{-224}  \\
& \nh \tablefootmark{d}& $2.1\pm 0.5 $ & $10^{22}$ \colc &\\
\hline
\multirow{2}{*}{Cold PION}&  $\log \xi$ \tablefootmark{c} & $-4.1\pm 0.1$ & \xiun
& \multirow{2}{*}{-358}  \\  
& \nh \tablefootmark{d} & $1.8\pm 0.8$ & $10^{22}$ \colc &\\
\hline
\end{tabular}
\tablefoot{
\tablefoottext{a}{Decrease of the C-statistic with respect to a model including
only the spectrum of the nucleus scattered by the PSF wings (C/Expected=430/45)}
\tablefoottext{b}{Plasma electron temperature.}
\tablefoottext{c}{Ionization parameter of the photoionized gas.}
\tablefoottext{d}{Hydrogen column density of the photoionized gas.}}
\end{table}
%
%
%
%
%
%
\section{Discussion}
\label{disc}
%
%
We present here an
analysis of a high-spatial-resolution Chandra ACIS-S
image of the bright quasar
1H 0419-577.  We
find that the source is
resolved for energies below 
$\sim$1 keV. Extended 
soft X-ray emission
is detected up to a distance
of at least $\sim$8\arcsec/16 kpc
from the nucleus. The 0.5--2.0
keV luminosity in the extended
emission is $\sim2 \times 10^{42}$
\ergs.
%
%
%
%
%
%
Our finding
confirms a previous detection
of extended emission
with ROSAT-HRI \citep{pre2001}.\\
%
%
We can certainly exclude
that the extended X-ray
emission is an artifact due to 
the scattering of the
source X-ray photons
by the dust in our Galaxy
\citep{pre1995}.
The foreground
dust reddening along the line of sight
of 1H 0419-577 (E(B-V)=0.015)
is similar to that of
the nearby Seyfert NGC 4051, for example.
This corresponds to 
a low \nh along our
line of sight and thus,
no evident X-ray scattering
halo is expected.
Indeed,
in the Chandra ACIS-S
image of NGC 4051,
no strong
extended emission 
is present
\citep{utt2003}.\\
%
%
Moreover, we can exclude
that we are observing a virialized
hot gas halo similar to those which
are often seen in early-type galaxies
\citep{for1979}.
For these hot gas halos the soft X-ray
luminosity
is tightly
correlated with the gas temperature
\citep{osu2003}.
For a gas temperature of $\sim$0.2 keV,
as we found here for 1H 0419-577, this
scaling relationship predicts a luminosity
of $\sim 9 \times 10^{39}$ \ergs.
This is at least
two orders of magnitude lower than the
luminosity that we find here for the
extended emission. \\
%
%
Thus, after having excluded
other possibilities,
the prime suspect for being
related with the
extended soft X-ray emission
is the galactic-scale [\ion{O}{iii}]
outflow, that is also
seen as a WA
in the X-ray/UV spectrum.
%
%
%
Galactic winds can be initially
driven either by a powerful
AGN disk wind \citep{kin2010} or by
the combination of stellar winds
and supernovae associated to
a starburst \citep{hec1990}. In the case
of 1H 0519-577,
the photoionization in the [\ion{O}{iii}]
outflow is certainly dominated by the AGN,
rather than by the SF.
Indeed, \citet{liu2015} find
that the ratio between the [\ion{O}{iii}] 
and the $\rm H \beta$ line is constantly above 10, 
up to a distance
of $\sim$5 kpc from the nucleus. This
line ratio indicates AGN dominance
according  to the
classification of \citet{bal1981}.
However,
in principle, the SF
may still be the driver of
the kinetic outflow. Using the relationship
given in \citet{rie2009} 
and the Spitzer-MIPS luminosity
at 24 $\mu \rm m$ \citep[$\sim9\times10^{10}$ \lsun,][]{spitzer},
we estimated a star formation rate
of $SFR \la$70 \msunyr for 1H 0419-577.
We consider the obtained value
to be an upper limit because the calculation
assumes that the infrared
luminosity is only due to the SF,
while in reality the nucleus (i.e., the
dusty torus surrounding the nucleus)
certainly contributes to a fraction
of it. The true SFR is likely
even lower than this. Indeed, in the near infrared,
the source displays an AGN-dominated
spectrum, lacking a prominent
polycyclic aromatic hydrocarbon (PAH)
feature at 3.3 $\rm \mu m$ 
\citep[$L_{\rm PAH \, 3.3 \mu m} \leq 4.5 \times 10^{8}$ \lsun,][]{yam2013}. 
These PAH molecules are considered a good
indicator of star formation activity
in AGN hosts \citep[e.g.,][]{shi2007}.\\
According to
the computations of  \citet{vei2005}, for example,
a constant $SFR$,
at a time t$\sim10^{7}$ yr,
can produce at most a wind kinetic
luminosity of $\sim7 \times 10^{41}$
\ergs for each solar mass of star formed.
Thus, in our case
this implies 
$L_{\rm kin} \la 5 \times 10^{43}$ \ergs, 
which is in principle  consistent with
the kinetic luminosity
estimated for the[\ion{O}{iii}] outflow
([0.01--7]$\times 10^{43}$ \ergs,
\citealt{liu2015}).
Hence, we cannot exclude
that the SF
drives or contributes in driving
the galactic outflow. We can,
however, exclude that the
X-ray luminosity of the
extended emission is due
to the combination
of high-mass X-ray binaries,
young supernova remnants,
and hot plasma associated
to the starburst. 
If that was the case, 
according to
the relationship of 
\citet{ran2003}, 
the expected soft X-ray luminosity
would be 
$L_{\rm 0.5-2.0\, keV}\la 3
\times 10^{41}$ \ergs.
This is at least one
order of magnitude lower
that the luminosity
that we measured for
the extended emission.\\
In our spectral analysis
we tested both a collisional-ionization
and
a photoionization scenario for
the physical origin of the extended
emission. 
Both models
predict emission
lines that could be, in principle,
individually resolved
in a higher-resolution spectrum.
We used the RGS spectrum
that we published in \citet{dig2013}
to check if the features
predicted by our models of
the extended emission are
consistent,
within the errors,
with higher-resolution
spectral data.  For this exercise
we recorded the luminosities
predicted by the CIE/PION model
for the most prominent
X-ray transitions
and we compare them with
the measured line luminosities
or upper limits derived
from the RGS data. We list
all the predicted and 
observed line luminosities
in Table \ref{lnrgs.tab}.\\
%
%
\begin{table}
\caption{Comparison between the predicted and the observed luminosities
of the emission lines of the extended emission component.}     
\label{lnrgs.tab}      
\centering                    
\begin{tabular}{l c c c c}        
\hline\hline                 
Ione & 
Energy 
& CIE lum \tablefootmark{a}
& PION lum\tablefootmark{a}
& RGS lum\tablefootmark{b}\\
& keV & log \ergs & log \ergs & log \ergs \\
\hline
\ion{C}{v} & 0.37 & 42.30 & & $\leq 41.55$\\
\ion{N}{ii} & 0.40 & & 42.65 & $\leq 41.04$\\
\hline

\hline
\ion{C}{vi} & 0.37 & 40.61 & 41.16 & $\leq 41.67$\\
\ion{N}{vi} & 0.42 & 39.72 & 40.41 & $\leq 40.73$\\
\ion{N}{vii} & 0.50 & 40.47 & 40.52 & $\leq 40.65$\\
\ion{O}{vii}-f & 0.56 & 41.26 & 41.42 & $41.84\pm0.08$\\
\ion{O}{viii} & 0.65 & 41.28 & 41.16 & $\leq 41.69$\\
\ion{Ne}{ix} & 0.90 & 40.58 & 40.80 & $\leq 40.52$\\
\hline
\end{tabular}
\tablefoot{
\tablefoottext{a}{Line luminosities predicted by the CIE/PION model for the extended emission.
Lines belonging to the cold and the warm component are listed in the upper and in the lower panel,
respectively.}
\tablefoottext{b}{Line luminosity or upper limit measured from the RGS spectrum
of \citet{dig2013}.}}
\end{table}
%
%

%
We found that the cold component PION/CIE
model predicts strong
emission lines from lowly ionized
carbon and nitrogen that
are not present in the data
(Table \ref{lnrgs.tab},
upper panel).
We note, however, 
that the ACIS effective area in the
energy range where
the cold PION/CIE
component dominates
the spectrum is affected by high uncertainty due to
a time-dependent 
molecular contamination
which has been building up
on the optical filters.
This is most likely
preventing us from achieving
a reliable parameterization
of the model in this energy
range. 
An observation using the
Low Energy Transmission
Grating Spectrometer
in combination
with the HRC detector
would be best suited to
studying the spectrum
in the low-energy range.
Conversely,
we found that
all the emission lines
predicted by the
warm component
of the PION/CIE model are
consistent with the 
RGS data (Table \ref{lnrgs.tab},
lower panel).
Thus, after this test,
both a collisionally ionized
and a 
photoionized plasma model
remain plausible explanations for the
spectrum of the extended emission
at energies comprised
between $\sim$0.5 and $\sim$1.0
keV.\\
In the case of  the collisionally
ionized model, the normalization
of the CIE model $n_{\rm e}n_{ \rm H}V$
(where $n_{\rm e}$ and $n_{\rm H}$ are the
electron and proton density, respectively,
and $V$ is the volume of the emitting
gas) provides an estimate
of the gas density. Assuming a filled
sphere with a radius of
16 kpc,
we obtain a gas number density
of the order of 
$10^{-3}\, \rm cm^{-3}$.
Such a low-density hot medium
may be a shocked gas component
associated to the galactic wind.
Analytical
models for AGN feedback
\citep{fau2012,kin2015} prescribe that
a powerful nuclear wind, 
initially driven by the
activity of the central supermassive
black hole, collides with the
surrounding medium and 
drives an outflow. Because
of the impact, a backwards
shock propagates in
the wind, while a forward
shock is driven into the surrounding
ISM. The shocked, expanding ISM is
expected to cool radiatively
\citep{zub2014,costa2015}
from temperatures of
the order of $10^{7}$ K 
down
to temperatures of the
order of $10^{4}$ K.
As the shock propagates
outward, free-free emission
becomes the prevailing cooling
mechanism and X-ray luminosities
comprised between $10^{41}$
and $10^{44}$ \ergs, 
where the exact value
depends mainly
on the location of
the shock radius
and on the density
of the ambient medium,
are predicted
\citep{nims2013}.\\
%
%
%
%
%
%
This interpretation
has also been proposed
for the case of 
SDSS J1356+1026
\citep{gre2014}
which is another quasar
where both a galactic-scale
[\ion{O}{iii}] outflow and
an extended soft X-ray
emitting nebula
were detected.
In this scenario,
the [\ion{O}{iii}]
and the X-ray/UV WA
may trace clouds of colder
swept-up material that are entrained
in the wind. A two-phase
wind structure, where
the line-emitting gas
is confined to colder
clumps that are in pressure
equilibrium with  a hot,
low-density outflow, has been
proposed for instance
in \citet{liu2013} to explain
the observed properties
of $\sim$kpc scale [\ion{O}{iii}]
outflow in radio-quiet quasars.
The warm photoionized gas may be in pressure
equilibrium with the hot medium
if $n_{\rm warm}T_{\rm warm} \sim n_{\rm hot}T_{\rm hot}$.
Using the parameters derived here for the
hot gas and taking $T_{\rm warm} \sim 10^{4} \rm K$ for
the photoionized gas, the pressure
equilibrium condition
prescribes $n_{\rm warm} \sim 0.1 \rm \, cm^{-3}$.
This is in principle
compatible with the upper limit
for the WA density derived
in \citet{edm2011} and with that
derived in \citet{liu2015} for the [\ion{O}{iii}]-emitting
material.\\
%
%
%
%
%
On the other hand,
we cannot rule out
the possibility that
we are instead observing
a nebula of photoionized gas. 
The AGN is able to photoionize
the gas up to a scale of a tenth
of a parsec, thus a photoionized
gas nebula extending up
to $\sim$16 kpc
is conceivable. 
In this scenario, a morphological
correspondence between 
the [\ion{O}{iii}]
and the soft X-ray emission
is expected. 
In our case,
a higher resolution
in the soft X-ray band would be needed
to access the true morphology
at scales comparable to the size
of [\ion{O}{iii}] lobes.
Nevertheless, we attempted
to perform a comparison
using a PSF deconvolved
image \citep[as in e.g.,][]{gom2017}
and, in this exercise,
we found that
the soft X-ray and
the [\ion{O}{iii}]
morphology do not match.
If confirmed, this lack
of morphological correspondence
could indicate that
the [\ion{O}{iii}] line and
the soft X-ray emission do no
not arise in the same 
photoionized medium.\\
%
%
%
%
%
%
In addition,
we found that the counterpart
in emission of the WA can account
only for a small fraction ($\sim2\%$)
of the extended emission luminosity.
Thus, the present analysis
confirms that a connection
between the line-emitting gas
and X-ray/UV WA
is unlikely in this case \citep{dig2013}.
Indeed, the global
parameters of the emitting
(i.e., $\log \nh \sim 22.3$ \colc
and $\log \xi \sim 1.3$ \xiun)
and of the absorbing
(i.e., $\log \nh \sim 19.9$ \colc
and $\log \xi \sim 0.03$ \xiun) gas
are largely inconsistent.
This implies
that the extended photoionized emitting gas must
have a low covering fraction
for absorption. Indeed, a photoionized
gas with the parameters found here
is expected to produce prominent absorption lines 
(e.g., \ion{O}{vii}-\ion{O}{viii},
Fe-UTA) in the
RGS band, that are not present in the data
In \citet{dig2013} we performed the
photoionization modeling of
the UV/X-ray narrow emission lines
in this source. In that exercise,
we found a covering fraction of 4\%
for the emitting gas. If we use this
value of covering factor
for the PION component found here,
the associated absorption lines are 
indeed consistent with
the Chandra-ACIS and the RGS data. Such
a low-covering fraction implies
that either the gas is fragmented
in small clouds or that it is mostly 
outside our line of sight. \citet{liu2015}
found that the rotational pattern suggested
by the median velocity of the [\ion{O}{iii}]
line is suggestive of disk galaxy
viewed almost edge-on ($\geq70\degr$). 
Moreover, in \citet{dig2014} 
we proposed that the AGN is seen
at high inclination.
In this geometry, in order not to intercept
our line of sight, the photoionized
soft X-ray emitting gas must be distributed mostly
along the polar directions, rather
than being spherically symmetrical.
If this was the case the UV/X-ray WA,
must trace a different gas component, 
physically disconnected from the emitting gas,
and located closer to the equatorial
plane.

\section{Summary and conclusion}
\label{sum}

We analyzed a Chandra-ACIS
high-spatial-resolution image 
of the Seyfert 1 galaxy 1H 0419-577.
This source is known for hosting
a galactic-scale outflow, which
is seen  as warm absorber
in the UV and in the X-ray spectrum,
and is also highlighted
in the [\ion{O}{iii}] image.\\
Extranuclear X-ray emission is
resolved by Chandra at energies
lower than 1 keV. The bulk 
of the soft X-ray emission
comes from a roughly 
spherically symmetrical
halo surrounding the nucleus. 
This extends up to a distance 
of $\sim$8\arcsec/16 kpc
and has a luminosity
of $\sim2 \times 10^{42}$ \ergs
in the 0.5--2.0 keV band.\\
The morphology of the soft X-ray extended
emission is spherically symmetric.
We were unable to recover
a morphological resemblance
between the soft X-ray extended
emission and the biconical outflow
that is traced by the 
[\ion{O}{iii}] emission line.\\
The analysis
of the spectrum of the extended emission
did not yield conclusive results about
its physical origin. We report that the
spectrum rises up sharply at energies
below $\sim$0.5 keV. However, due to
the calibration uncertainty of the ACIS 
effective area in this energy range, we 
are unable to find an accurate
parameterization of any model
in this spectral range. \\
In the
energy range between 0.5 end 1 keV, 
both a collisionally ionized
($T_{\rm e}=0.22\pm0.02$ keV)
and a photoionized plasma-model 
($\log \xi=1.3\pm0.1$
$\nh=2.1\pm 0.5 \times 10^{22}$ \colc)
best-fit
the ACIS spectrum and are consistent with
the RGS spectrum already published 
in \citet{dig2013}.  \\
In the former case,
we may be observing a thermally emitting
gas bubble, which was inflated by the
wind shock and that is radiatively cooling
down. In this scenario the [\ion{O}{iii}]
emission line and the X-ray/UV warm absorber
may trace cooler clumps that are entrained
in the hot wind. \\
Alternatively, we may
be observing a large-scale photoionized
gas nebula. In this case,
we can exclude that
the extended photoionized gas
is the counterpart in emission
of the X-ray/UV warm absorber.
Thus, we infer that the extended
photoionized gas nebula must have
a low covering fraction for absorption.
This implies that either the gas
is fragmented or that it is
distributed mostly along the
polar directions, outside
our line of sight.\\ 

%

\begin{acknowledgements}

The scientific results reported in this article are based 
on observations made by the Chandra X-ray Observatory.
LDG acknowledges support from the Swiss National Science Foundation.
SRON is financially supported by NWO, the Netherlands organization
for Scientific Research. We thank Guilin Liu for kindly providing the GEMINI-IFU image, and Francesco Tombesi for useful discussions.

\end{acknowledgements}

%
%
%
%
%
%
%

\end{document}